\documentclass{nature}

\usepackage{lineno}
\usepackage{lmodern}
\usepackage{amsmath}
\usepackage{amssymb}
\usepackage{graphicx}
\usepackage{hyperref}
\usepackage{xspace}
\usepackage{caption}
\usepackage{mathptmx}
\usepackage{mathrsfs}
\usepackage{pdfpages}
\usepackage{mathtools} 
\usepackage{bm}

\usepackage{xr-hyper} 
\title{Ultrafast current and field driven domain-wall dynamics in van der Waals antiferromagnet MnPS$_3$ 
} 

\author{Ignacio M. Alliati$^{1}$, Richard F. L. Evans$^2$, Kostya S. Novoselov$^{3,4}$, \& Elton J. G. Santos$^{5,\dagger}$}

\makeatletter
\let\saved@includegraphics\includegraphics
\AtBeginDocument{\let\includegraphics\saved@includegraphics}
\renewenvironment*{figure}{\@float{figure}}{\end@float}
\makeatother

\begin{document}

\maketitle

\begin{affiliations}
\item School of Mathematics and Physics, Queen's University Belfast, BT7 1NN, UK
\item Department of Physics, The University of York, YO10 5DD, UK 
\item Department of Material Science \& Engineering, National University of Singapore, Block EA,  
9 Engineering Drive 1, 117575, Singapore 
\item Chongqing 2D Materials Institute, Liangjiang New Area, Chongqing 400714, China 
\item Institute for Condensed Matter Physics and Complex Systems, School of Physics and Astronomy, The University of Edinburgh, EH9 3FD, UK \\
$^{\dagger}$email:   esantos@ed.ac.uk  
\end{affiliations}
 
\date{}

\begin{abstract}
The discovery of magnetism in two-dimensional (2D) van der Waals (vdW)
materials\cite{firstCrI3,CrGeTe,Guguchiaeaat3672,Klein1218} has flourished 
a new endeavour of fundamental problems in magnetism 
as well as potential applications in computing, sensing and 
storage technologies\cite{Novoselov19, Gongeaav4450,Li:2018aa,Santos:2013aa,Hong:2019aa,Tian:2020aa}. 
Of particular interest are antiferromagnets 
\cite{MnPS3-1986,Makimura93}, 
which due to their intrinsic antiferromagnetic exchange 
coupling show several advantages in relation to ferromagnets such as  
robustness against external magnetic perturbations. 
This property is one of the cornerstones of 
antiferromagnets\cite{RevModPhys.90.015005} 
and implies that information stored in 
antiferromagnetic domains is invisible 
to applied magnetic fields preventing it from 
being erased or manipulated. 
Here we show that, despite this fundamental understanding,
the magnetic domains of recently discovered 
2D vdW MnPS$_3$ antiferromagnet\cite{Long:2017aa,Kim2019}  
can be controlled via external magnetic fields and electric currents. 
We realize ultrafast domain-wall dynamics with velocities up to 
$\sim$1500 m s$^{-1}$ and $\sim$3000 m s$^{-1}$ respectively to a 
broad range of field magnitudes (0.0001$-$22 T) and 
current densities ($10^{8}-10^{10}$ A cm$^{-2}$). 
Both domain wall dynamics are determined by the 
edge terminations which generated uncompensated spins
following the underlying symmetry of the 
honeycomb structure. 
We find that edge atoms belonging to different magnetic  
sublattices function 
as geometrical constrictions preventing the 
displacement of the wall, whereas having 
atoms of the same sublattice at both edges of the material allows for 
the field-driven domain wall motion which is only 
limited by the spin-flop transition of the antiferromagnet 
beyond 25 T. 
Conversely, electric currents 
can induce motion of domain walls in most of the edges 
except those where the two sublattices are present at the 
borders (e.g. armchair edges). Furthermore, the orientation of the layer 
relative to the current flow 
provides an additional degree of freedom 
for controlling and manipulating magnetic domains in MnPS$_3$. 
Our results indicate that the implementation of 
2D vdW antiferromagnets in real applications requires  
the engineering of the layer edges which enables 
an unprecedented functional feature in ultrathin device platforms. 
\end{abstract}

\noindent 


The emergence of magnetism in 2D vdW materials has opened exciting new avenues 
in the exploration of spin-based applications at the ultimate level of few-atom-thick layers. 
Remarkable properties including giant tunnelling 
magnetoresistance\cite{Song1214,Wang2018-hq,Klein1218} and 
layer stacking dependent magnetic phase\cite{Song:2019aa,Xiaodong19} 
have recently been demonstrated.  Even though these studies show that 
rich physical phenomena can be observed in 2D ferromagnets, 
the dynamics of domain walls which determine whether 
such compounds can be effectively implemented in real life 
device platforms remains elusive. Very few 
reports have shed some light 
on the intriguing behaviour of magnetic domains\cite{Maletinsky19,Zhong:2020aa}  
and their walls\cite{Wahab20} in ferromagnetic 
layered materials. 
The scenario is even less clear for 2D antiferromagnets where the 
antiferromagnetic exchange coupling between spins adds a level of complexity 
in terms of the manipulation of the magnetic moments 
by conventional techniques as zero net magnetization 
is obtained\cite{RevModPhys.90.015005}. 
Indeed, recent measurements using 
tunnelling magnetoresistance, a common 
approach for ferromagnetic materials, unveiled that 
antiferromagnetic correlations persist down to the level of 
individual monolayers of MnPS$_3$\cite{Long:2020aa}. 
This result suggests that yet unexplored ingredients at  
low dimensionality play an important role in the detection 
and manipulation of the antiferromagnetic order in 2D vdW compounds. 
Moreover, how domain walls in MnPS$_3$ behave 
and can be controlled externally in functional devices for 
practical applications are still open questions. 

Here we show that 
electrical currents and, unexpectedly, magnetic fields can move 
domain walls in monolayer MnPS$_3$ at low-temperatures achieving
fast velocities within the km s$^{-1}$ limit. 
While bulk antiferromagnetic compounds are insensitive 
to magnetic fields, the interplay between low-dimensionality and 
edge-type offers control over domain wall dynamics 
via an initially unthinkable external parameter. 
In configurations where the layer terminates with either a zigzag array 
of Mn atoms or dangling-bonds, the domain walls are controllable via 
both currents and magnetic fields at a broad range of magnitudes. 
For configurations where the edge atoms assume an 
armchair configuration, the domain wall appears pinned  %
and no motion is observed irrespective of the 
field intensity or current density applied. 
Our results indicate a rich variety of possibilities depending on the 
edge roughness and introduces the layer termination as one of 
the determinant factors for integration of 
2D antiferromagnets in novel domain-wall based applications.  


We firstly investigate how magnetic domains are formed in monolayer 
MnPS$_3$ through simulating the zero-field cooling process 
for a large square flake of 0.3 $\mu$m $\times$ 0.3 $\mu$m 
using atomistic spin dynamics which incorporate atomistic (several \AA's)
and micromagnetic ($\mu$m-level) underlying details 
(see full set of details in Supplementary Notes \ref{SI-methods-dft}$-$\ref{SI-LLG1}). 
The system is thermally equilibrated above the Curie temperature at 80 K
and then linearly cooled to 0 K in a simulated time of 4.0 ns as shown 
in Figure \ref{fig1} and Supplementary Movies S1-S2. 
The time evolution of the easy-axis component of the 
magnetization S$_{\rm z}$ is used to display the nucleation of the 
magnetic domains at different temperatures and magnetic fields. 
While domain walls appeared at zero field 
with a large extension over the 
simulation area (Fig. \ref{fig1}{\bf a-e}), an external field 
can flush out any domains resulting in a homogeneous magnetization 
after 2 ns (Fig. \ref{fig1}{\bf f-j}). We also observed that some simulations 
at zero field ended up in the formation of a monodomain throughout the surface.
This suggests that antiferromagnetic domains are not intrinsically 
stable in MnPS$_3$ similarly as in ferromagnetic layered materials, e.g. CrI$_3$\cite{Wahab20}. 
Indeed, the metastability of the domains 
prevents the wall profiles from reaching a truly ground-state configuration 
as they initially appears {winded} with several nodes at 0 K (Fig. \ref{fig1}{\bf d}), 
but incidentally evolved to an unwound state (Fig. \ref{fig1}{\bf e}). 
A close look reveals a continuous rotation of the spins 
over the extension of the wall pushing the nodes out 
of the domain wall profile (Supplementary Figure \ref{SI-wind-DW} and Supplementary Movie S3). 
Such interactions between nodes can extend as long as $\sim$95 nm along the wall which is 
more that two order of magnitudes larger than the thickness (0.8 nm\cite{Long:2017aa}) 
of the monolayer MnPS$_3$. At longer times, the domain wall reaches stability and 
does not show any sudden variations on the spin configurations. 

To determine whether the interplay between metastability and the high 
magnetic anisotropy of MnPS$_3$ could give additional features to the domain walls, 
we analyse the local behaviour of the spins in the domain wall (Figure \ref{fig2}). 
We notice that as the spins rotate from one magnetic domain to 
another they tend to align with the 
zig-zag crystallographic direction displaying an angle  
of $\phi=64.02^{\rm o}$ (Fig. \ref{fig2}{\bf a-b}). 
The projections of the 
total magnetization at the wall over the out-of-plane (S$_{\rm z}$) and 
in-plane (S$_{\rm x}$, S$_{\rm y}$) components  
show sizeable magnitudes of S$_{\rm y}$ and S$_{\rm x}$ as 
the spins transition from one domain to another despite the 
easy-axis anisotropy along of S$_{\rm z}$ (Fig. \ref{fig2}{\bf c-d}). 
This indicates a domain wall of hybrid characteristics rather than one of 
Bloch and N\'eel type (Fig. \ref{fig2}{\bf e-f}).   
We can extract the domain wall width $\sigma_{\rm x,y,z}$ by fitting 
the different components of the magnetization 
(S$_{\rm x}$, S$_{\rm y}$, S$_{\rm z}$) 
to standard equations\cite{Hubert-book} of the form: 
%
\begin{eqnarray}
S_{j} &=& \frac{1}{\cosh(\pi (j-j_0)/\sigma_j)},  ~~~{\rm with} ~~~j=x, y  \label{eq:Sxy}\\ 
S_{z}&=& {\tanh} (\pi (z-z_0)/\sigma_z)   \label{eq:Sz}
\end{eqnarray}
where $j_0$ and z$_0$ are the domain wall positions at in-plane 
and out-of-plane coordinates, respectively. The domain wall 
widths are within the range of $\sigma_{\rm x,y,z}=3.40-3.50$ nm.  
Such small widths are commonly observed in permanent magnetic 
materials\cite{Hubert-book} due to their exceptionally high magnetic 
anisotropy such as Nd$_2$Fe$_{14}$B (3.9 nm), SmCo$_5$ (2.6 nm), 
CoPt (4.5 nm) and Mn overlayers on Fe(001) (4.55 nm)\cite{Kirschner04}. 
In these systems, magnetic domains are 
energetically stable after zero-field cooling due to 
long range dipole interactions which were also checked in our study resulting 
in no modifications of the results. 
Therefore, MnPS$_3$ reunites characteristics 
from soft-magnets (large area uniform 
magnetization) and hard-magnets 
(large magnetic anisotropy, narrow domain walls) 
within the same material.

An outstanding question raised by these hybrid features is whether the 
domain walls can be manipulated by electric currents and magnetic fields. 
It is well known that antiferromagnets are insensitive 
to external magnetic fields but are rather controllable through currents particularly 
in high-anisotropy materials\cite{RevModPhys.90.015005,Jungwirth:2016aa}.  
However, the low dimensionality together 
with the underlying symmetry of the honeycomb structure may lead 
to novel features on the dynamics of domain walls not yet observed 
in bulk antiferromagnets. To investigate this we simulate 
the spin-transfer torque induced by 
spin-polarized currents and the effect of magnetic fields on a 
large nano-flake of monolayer MnPS$_3$ of dimensions of 300 nm$\times$50 nm
(see Supplementary Note \ref{SI-torque} for details). 
The domain wall is initially stabilized from one-atom-thick wall 
which broadens and develops a profile during the 
thermalization over a simulation time of 0.5 ns. The system is then allowed to 
evolve for longer times ($\sim2$ ns) to ensure that 
no changes are observed in the system close to the end of the dynamics. 

Surprisingly, both electric currents and magnetic fields are able to induce 
the motion of domain walls in the antiferromagnetic MnPS$_3$ 
resulting in a broad range of velocities (Figure \ref{fig3}). 
For current-induced domain wall motion, 
wall velocities up to $v=$3000 m s$^{-1}$ 
are seen at a maximum current density of 
$j=80 \times 10^{9}$ A cm$^{-2}$ (Fig. \ref{fig3}{\bf a-b}). 
At such large values of $j$, we observe primarily 
two regimes that are characterized by 
different dependences of $v$ with $j$. 
For $j \leq 30 \times$ 10$^{9}$ A cm$^{-2}$ (Fig. \ref{fig3}{\bf b}) 
a linear dependence  
is noticed which can be described 
by a one-dimensional model (Supplementary Note \ref{SI-sec:1D-model}) as: 
\begin{equation} \label{eq:linear_v}
v= C_{c} j 
\end{equation}
where $C_{c}=\frac{\mu_{B} \sigma}{2 \alpha e m_{s} t_{z}}\theta_{SH}$, with $\mu_{b}$ 
the Bohr magneton, $\sigma$ the domain wall width, 
$\theta_{SH}$ the spin Hall angle, 
$\alpha$ the Gilbert damping parameter, $e$ the electron charge, 
$m_{s}$ the modulus of the magnetization per lattice, and $t_{z}$ the layer thickness. 
Eq.\ref{eq:linear_v} is consistent with adiabatic spin-transfer 
mechanisms in thin 
antiferromagnets\cite{RevModPhys.90.015005,Jungwirth:2016aa,PhysRevLett.117.087203} 
where the conduction electrons from the 
current transfer angular momentum 
to the spins of the wall which keeps its 
coherence through a steady motion (Fig. \ref{fig3}{\bf a} and 
Supplementary Movie S4). 
The value of $C_{c} =67.11 \times 10^{-13}~m^{3} {C}^{-1}$ 
extracted from our simulation data helps to 
find other parameters not easily accessible in experiments or 
from theory, e.g. $\theta_{SH}$. The magnitude of $\theta_{SH}$
determines the conversion efficiency 
between charge and spin currents, and it is the figure of merit 
of any spintronic application.
Using the definition of $C_{c}$ (Supplementary Note \ref{SI-sec:1D-model}), 
we can estimate $\theta_{SH}(\%)= 0.010$ which is 
comparable to standard heterostructures and 
antiferromagnets\cite{RevModPhys.90.015005} but at a 
much thinner limit. This suggests MnPS$_3$ as a potential 
layered compound for power-efficient device platforms. 
For $j \geq 40 \times$ 10$^{9}$ A cm$^{-2}$
the wall velocities tend to saturate to a maximum magnitude near 
3000 m s$^{-1}$ (Fig. \ref{fig3}{\bf b}) with a deviation from the linear dependence 
observed previously (Eq. \ref{eq:linear_v}). 
This intriguing behaviour can be understood in terms of the 
relativistic kinematics of antiferromagnets\cite{Haldane83,RevModPhys.90.015005}.
As the wall velocities approach the maximum group velocities ($v_{g1}$, $v_{g2}$), 
which sets the maximum speed for spin interactions into the system, 
relativistic effects in terms of the Lorenz invariance become more predominant. 
This is due to the finite inertial mass of the antiferromagnetic 
domain wall which can be decomposed into spin-waves 
represented through relativistic wave 
equations\cite{Haldane83,PhysRevB.90.104406,PhysRevLett.117.087203}. 
We can extend this idea further in a 2D vdW antiferromagnet  
by examining the variations of several quantities 
via special relativity concepts. For instance, 
the variation of the wall velocities versus current 
densities can be well analysed using: 
\begin{equation}\label{eq:relativ}
v= C_{c} j \sqrt{1-(v/{v_{g2}})^2}
\end{equation} 
where $v_{g2}$ is the maximum spin-wave group velocity 
at one of the branches of the magnon dispersion 
(Supplementary Note \ref{SI-sec:magnon}). Eq.\ref{eq:relativ} 
includes quasi-relativistic corrections\cite{PhysRevB.90.104406,PhysRevLett.117.087203} 
to the linear dependence recorded 
at low values of the density (Eq. \ref{eq:linear_v})
and can be solved self-consistently in $v$ for each 
magnitude of $j$. 
Strikingly, Eq.\ref{eq:relativ} provides 
an accurate description of the wall velocity
not only at low magnitudes of the density, where relativistic effects are 
rather small, but also for $j \geq 40 \times$ 10$^{9}$ A cm$^{-2}$. 
At such limit, the domain wall width $\sigma$ and 
the domain wall mass M$_{\rm DW}$ also 
shrinks and expands, respectively, 
exhibiting effects similar to the Lorentz contraction (Fig. \ref{fig3}{\bf d-e}). 
These phenomena can be reasoned by (see Supplementary Note \ref{SI-sec:1D-model} for details): 
\begin{eqnarray} 
 \sigma &=& \sigma_{o} \sqrt{1-(v_{}/v_{g2})^2}  \label{eq:width}   \\                 
   {\rm M}_{\rm DW}&=&\frac{2 \rho w t_z \pi}{\sigma_o\sqrt{1-(v_{}/v_{g2})^2}}   \label{eq:mass}
\end{eqnarray}
where $\sigma_{o}$ is domain wall width at low-velocities ($\sim$3.41 nm), 
$\rho=\frac{1}{J_{1NN} \gamma^2}$ (with $J_{1NN}$ the exchange parameter for the 
first nearest-neighbours and $\gamma$ the gyromagnetic ratio),  
$w$ is the width of the stripe of the material, and $t_z$ is the layer thickness. 
There is a sound agreement between the simulation data (Fig. \ref{fig3}{\bf d-e}) and 
Eqs. \ref{eq:width}$-$\ref{eq:mass} over a wide range of 
velocities with minor deviations occurring 
above 2500 m s$^{-1}$ due to non-linear spin excitations. 
We found that at such large wall velocities, 
spin waves or magnons are emitted throughout the layer with frequencies 
in the terahertz regime (Fig. \ref{fig3}{\bf f-g}). 
%
%
These excitations can be found in the wake of the wall 
forming simultaneously in front and behind  
the wall motion (see Supplementary Movie S5).  
Analysing the variations of S$_{\rm z}$ over time 
at different $j$ (Supplementary Figure \ref{SI-turbulent-DW}), 
we noticed that high currents 
generated precession of the spins around the easy-axis 
with their high in-plane projections (S$_{\rm x}$, S$_{\rm y}$) being 
transmitted through spin-waves into the system (Fig. \ref{fig3}{\bf f-g}). 
This is critical at large values of $j$ where the 
variation of the position of the domain wall with time 
results in two velocities before and after the 
magnons start being excited (Supplementary Figure \ref{SI-turbulent-DW}).  
This behaviour is particularly turbulent at longer 
times as the wall profile can not be defined any more 
with the appearance of several vortex, antivortex and 
spin textures at both edges of the layer 
and inside the flake (Supplementary Movie S6). 


To have a deeper understanding of the characteristics 
of spin excitations on the domain wall dynamics in MnPS$_3$, 
we have developed an analytical model 
using linear spin-wave theory\cite{PhysRev.117.117,PhysRev.87.568} 
that accounts on the magnon dispersion $\varepsilon(\mathbf{k}$)  
and their group velocities 
$v_{g}(\mathbf{k})=\frac{\partial \varepsilon(\mathbf{k})}{\partial \mathbf{k}}$ 
over the entire Brillouin zone. 
Supplementary Note \ref{SI-sec:magnon} 
provides a full description of the details involved.  
The maximum spin-wave group velocities 
($v_{g1}$, $v_{g2}$) that MnPS$_3$ can sustain at different magnon 
branches (Supplementary Figure \ref{SI-group-vel}) 
are in the range from $v_{g1}=2323$ m s$^{-1}$ 
to $v_{g2}=3421$ m s$^{-1}$ (Fig. \ref{fig3}{\bf b}).
These magnitudes correspond to the highest velocities 
at which spin excitations can propagate in the system and 
put the lower ($v_{g1}$) and upper limits ($v_{g2}$) where 
magnons participate in the domain-wall dynamics. 
Indeed, there is a good agreement with the numerically 
calculated wall-velocities where 
the spin-waves start being emitted into the sheet   
($\sim$2248 m s$^{-1}$), and the wall saturates to its 
maximum speed ($\sim$2970 m s$^{-1}$). 
The slightly lower values obtained in the simulations relative to 
$v_{g1}$ and $v_{g2}$ are due to 
the effect of damping on the propagation of domain 
walls due to the emission of spin-waves.  
A similar feature has been observed in the past 
in 3D ferromagnetic\cite{PhysRevB.81.024405,PhysRevLett.65.2587} 
and antiferromagnetic\cite{PhysRevLett.117.087203,PhysRevLett.112.147204} 
compounds but the emergence of such 
phenomena in a 2D vdW antiferromagnet is unprecedented. 
Additionally, we can estimate a maximum wave frequency 
($f_{max}=\hbar \varepsilon/2\pi$) corresponding to $v_{g2}$ 
of about 4.03 THz. This value surpasses those 
measured in the state-of-the-art antiferromagnetic materials such as 
in MnO\cite{Nishitani:2013aa}, NiO\cite{Kampfrath:2011aa,PhysRevLett.105.077402}, 
DyFeO$_3$\cite{Kimel:2005aa}, HoFeO$_3$\cite{Mukai:2014aa} 
and heterostructures combining 
MnF$_2$ and platinum\cite{Vaidya:2020aa} by several times. 
This implies that antiferromagnetic domain walls in MnPS$_3$ 
can be used as a terahertz source of electric signal at 
the ultimate limit of a few atoms thick layer.

Remarkably, the application of a magnetic field results  
in a very counter-intuitive behaviour  
as the domain wall moves with 
velocities as high as $\sim$1500 m s$^{-1}$ (Fig. \ref{fig3}{\bf h-j} 
and Supplementary Note \ref{SI-sec:field_dw} 
for additional discussions).  
We can fit most of the field-induced domain wall dynamics for $B\leq 20$~T with: 
\begin{equation}\label{eq:v_B}
v=86.28 ~B 
\end{equation}
with a linear regression coefficient of $R^{2}=9996$. 
The motion is steady, keeping the wall 
shape throughout the motion. We observe however that both the domain wall width and the 
domain wall mass change their magnitudes in opposite trend as that observed in the 
current-driven domain wall dynamics (Supplementary Figure \ref{SI-width_mass_B}). 
We attribute this difference to the distinct operation of the 
external stimulus on the domain wall. In the current driven case, 
the action is tightly focused at the centre of the wall, where the angular 
change in neighbouring spins is the largest. For large currents the wall is 
not able to relax fully leading to relativistic contraction in the wall width. 
In contrast, the magnetic field acts across the whole wall and tends to 
strengthen the spin flop (SF) state, which in turn leads to an 
increase in the domain wall width with increasing field 
strength (Supplementary Figure \ref{SI-width_mass_B}{\bf a}).
This effect is sufficient to counteract the relativistic effect (which is also present) 
and leads to a net decrease of the domain wall 
mass (Supplementary Figure \ref{SI-width_mass_B}{\bf b}). 
Wider domain walls naturally have lower mass as they are easier to move until 
SF states are achieved for fields above 22 T.  
In addition, some curvature is formed as the wall moves 
with its starting points from the terminations of the sheet 
parallel to the wall movement (Fig. \ref{fig3}{\bf a, h} and 
Supplementary Movie 7). The spins around one edge move in 
advance relative to those at the middle of the system and at 
the opposite edge creating a curved wall during the 
motion (see detailed features in Supplementary Movie 8). 
Such deviation from the planar wall shape has been 
reported in hetero-interfaces formed by 
NiFe/FeMn bilayers\cite{Chien00} 
but not yet in a monolayer of a 2D vdW antiferromagnet. 
This indicates a direct relation between domain-wall 
motion and the material geometry via edge roughness 
similarly as in magnetic wires\cite{Nakatani:2003aa}. 
A close look unveils that the type of edge 
plays a pivotal role in the 
domain wall dynamics induced by both 
magnetic fields and electric currents.
Sheets terminated with edge atoms in zig-zag (ZZ) and 
dangling-bond (DB) configurations (Fig. \ref{fig3}{\bf h}) 
in any combination (e.g. ZZ-ZZ, ZZ-DB, DB-DB) 
are susceptible to be manipulated by currents. 
Nevertheless, only domain wall in layers with dissimilar edges (e.g. ZZ-DB) can be 
controlled by magnetic fields. Borders formed by 
atoms in the armchair (ARM) configuration remain inert 
irrespective of the stimulus applied (Supplementary Figure \ref{SI-arm1}). 
Intriguingly, ARM edges under applied currents 
show a short displacement of the domain wall at 
earlier stages of the dynamics ($\sim$0.07 ns) but rapidly 
stabilizes to a constant position at longer times. 
As the current flows through the wall, the spins feel 
the torque induced by the 
spin-polarized electrons but rather than reorient 
the spins to follow the current direction, 
the spins at the wall precess around the easy-axis 
with no motion of the domain-wall. 
This mechanism is 
shown in details in Supplementary Movies S9-S10. 
The ARM edge in this case works as an effective 
pinning barrier for domain-wall propagation. 

The control of the domain-wall motion in an 
antiferromagnetic material via magnetic fields opens a new ground in the 
investigation of the role of edges on 2D magnetic materials. The 
fundamental ingredient that enables such phenomena is based on 
the underlying magnetic sublattices (e.g. A or B) composing the 
honeycomb structure (Fig. \ref{fig4}{\bf a-c}). 
Despite the border considered, for edge atoms residing at different sublattices 
the magnetic field induces  a torque at each sublattice that mutually compensates 
each other generating no net displacement of the domain-wall (Supplementary Movie S11). 
For edge atoms at the same sublattice the effect is additive inducing the translation 
of the wall. Indeed, we can further confirm this mechanism analysing the 
spin interactions present in the system on a basis of a generalized XXZ 
Heisenberg Hamiltonian in the form of: 
\begin{equation}
\mathscr{H} =- \sum_{\langle i,j \rangle} J_{ij} {\bf S}_i \cdot {\bf S}_{j} - \sum_{\langle i,j \rangle} \lambda_{ij} S_i^z S_j^z - D \sum_{i} \left(S_i^z\right)^2 - \mu_s \sum_{i} {\bf S}_i \cdot {\bf B}_i
\label{BfieldHamilton}
\end{equation}
where $J_{i,j}$ is the bilinear exchange interactions 
between spins {\bf S}$_i$ and {\bf S}$_j$ at sites $i$ and $j$, 
$\lambda_{ij}$ is the anisotropic exchange,
$D$ is the on-site magnetic anisotropy 
and {\bf B}$_i$ is the external magnetic field applied along the easy-axis (e.g.~B$_{\rm z}$). 
We only include bilinear exchange terms in Eq. \ref{BfieldHamilton} 
since biquadratic exchange interactions 
are negligible in MnPS$_3$\cite{Kartsev19}. 
We considered pair-wise interactions in 
$\langle i,j \rangle$ up to the third nearest 
neighbours ($3NN$) (Supplementary Table \ref{SI-tbl:table2}). 
All parameters are calculated using strongly 
correlated density functional theory based on 
Hubbard-$U$ methods. Supplementary 
Notes \ref{SI-methods-dft}$-$\ref{SI-mag-parameters} convey 
the full details of the approaches employed. 
Eq. \ref{BfieldHamilton} is then applied to calculate 
the spin interactions into the system taking into 
account any angular variations $\theta$ of the 
spins induced by the field.  
We determine the stability of the system before 
and after the application of B$_{\rm z}$ distinguishing  
the atoms away from the domain wall from 
those at the wall (Fig. \ref{fig4}{\bf d}). 
Such procedure is instrumental to unveil the influence 
of the edges on the energetics of the domain-wall dynamics 
as the atoms at these two spatial regions may respond differently to a 
magnetic perturbation. In fact, we found that spins 
that are distanced from the domain-wall (i.e. spin-up for $\theta=0$, 
and spin-down for $\theta=180^{\rm o}$) do not 
suffer any angular variation with B$_{\rm z}$ 
as the layer reached a new ground-state. 
Nevertheless, for spins at the domain-wall the 
new ground-state under a finite field (B$_{\rm z}\neq0$) is obtained at a 
value of $\theta$ different to that at B$_{\rm z}=0$ (Fig. \ref{fig4}{\bf d-f}). 
This indicates that the wall spins tend to rotate under magnetic 
fields and the effect is particularly strong for atoms at the 
edges. The variations in energy $-\Delta E$ 
show that when the two edges are 
similar (Fig. \ref{fig4}{\bf e}) 
two different sublattices will be localized 
at the borders which will respond likewise 
generating similar variation of energies.  As the 
atoms at the edges are more uncoordinated relative to those in 
the bulk of the system, they gain more energy from aligning with  
B$_{\rm z}$ which allows the spins to rotate 
more freely but in opposite direction compensating 
any displacement of the domain-wall. 
The scenario is completely different when the atoms at the edges belong 
to the same sublattice such as in a ZZ-DB layer (Fig. \ref{fig4}{\bf f}). 
In this case, $-\Delta E$ has a larger variation at the DB edge due to the lesser 
coordination with neighbouring atoms and consequently less exchange energy. 
This results in a more prompt rotation of the spins at the DB edge than that at 
the ZZ edge dragging the wall slightly ahead with the field (Supplementary Movie S8). 
Even though our analysis has been applied for MnPS$_3$ 
it should be universal for any antiferromagnetic layered material 
with a honeycomb lattice.

One of the main implications for having uncompensated spins at the 
edges selectively 
controlling domain-wall motion in 2D vdW antiferromagnets is that 
depending how the layer is oriented in a device-platform 
we can have many possibilities to induce domain wall dynamics. 
By engineering the type of edges in MnPS$_3$ we can either induce 
a fast domain-wall dynamics through both current and magnetic fields, 
or no motion whatsoever via geometrical constrictions. 
Our findings suggest that 2D layered antiferromagnets 
would not be invisible to common magnetic probes 
(e.g. tunnelling magneto-resistance\cite{Long:2020aa}) 
which is normally problematic 
for materials that hold antiferromagnetic coupling between the spins 
within the layer. In this case the antiferromagnetic layer 
can play a more active role in magnetic structures rather than induce exchange bias 
in an adjacent ferromagnetic layer\cite{RevModPhys.90.015005}. 
With the rapid integration of magnetic layered materials 
in applications and the discovery of more compounds 
with similar characteristics, our predictions will open novel
grounds in the investigations of edge-mediated 2D 
antiferromagnetic spintronics

\section*{Supplementary Materials}
\label{sec:org3881bef}

Supplementary Notes 1$-$15, Supplementary Movies 1$-$13 and 
Supplementary Figures 1$-$16.


\subsubsection*{Data Availability} 

The data that support the findings of this study 
are available within the paper and its Supplementary Information.  

\subsubsection*{Competing interests}
The Authors declare no conflict of interests.

\subsubsection*{Acknowledgments}
RFLE gratefully acknowledges the financial support of 
ARCHER UK National Supercomputing Service via 
the embedded CSE programme (ecse1307). 
EJGS acknowledges computational resources through the 
UK Materials and Molecular Modelling Hub for access to THOMAS supercluster, 
which is partially funded by EPSRC (EP/P020194/1); CIRRUS Tier-2 HPC 
Service (ec131 Cirrus Project) at EPCC (http://www.cirrus.ac.uk) funded 
by the University of Edinburgh and EPSRC (EP/P020267/1); 
ARCHER UK National Supercomputing Service (http://www.archer.ac.uk) via 
Project d429. EJGS acknowledges the 
EPSRC Early Career Fellowship (EP/T021578/1) and 
the University of Edinburgh for funding support. 

\subsubsection*{Author Contributions} 
EJGS conceived the idea and supervised the project. 
IMA performed ab initio and Monte Carlo simulations 
under the supervision of EJGS. IMA and EJGS elaborated 
the analysis and figures. RFLE implemented the spin-transfer-torque 
method. EJGS wrote the paper with inputs from all authors.  
KSN helped in the analysis and discussions. 
All authors contributed to this work, 
read the manuscript, discussed the results, and agreed 
to the contents of the manuscript.

\section*{References and Notes} 
\bibliography{references}     

\pagebreak 


\subsubsection*{Figure captions}

\begin{figure}[htbp]
\centering
\caption{\label{fig1}\textbf{Magnetic domain evolution of a 2D antiferromagnet.} 
Snapshots of the dynamic spin configuration of monolayer MnPS$_3$ 
during field-cooling at different temperatures (K), time steps (ns) and 
different magnetic fields: {\bf a-e,} 0 T and {\bf f-j} 0.2 T. 
The out-of-plane component of the magnetization S$_{\rm z}$ is 
used to follow the evolution in a 0.3 $\mu$m$\times$ 0.3 $\mu$m square flake. 
Labels on temperatures and time are the same for both 
magnitudes of the field at the same column. 
Color scale in {\bf e} shows the 
variation of S$_{\rm z}$. To provide a better visualization of the domains 
we inverted the colour scheme for the two sublattices. That is, 
spin up (spin down) corresponds to red (blue) 
for one of the sublattices, and blue (red) for the other. This convention 
results in a single colour for a given domain. 
See Supplementary Note \ref{SI-domain-colors} for further details. 
}
\end{figure}

\begin{figure}[htbp]
\centering 
\caption{\label{fig2} {\bf Hybrid domain wall formation and spin rotation.} 
{\bf a-b,} Local and global view, respectively, of a snapshot of one of 
the spin configurations in a 300 nm $\times$ 50 nm ribbon of  
MnPS$_3$. 
The small rectangle in {\bf b} corresponds to area studied in {\bf a}. The out-of-plane  
component of the magnetization S$_z$(color map) is utilized to monitor 
the formation of the domain wall. Spins rotated 
across the wall in pairs forming an angle $\phi$ 
with the zig-zag crystallographic direction of the 
honeycomb lattice of 64.02$^{\rm o}$. The system is at zero magnetic field and 0 K. 
{\bf c-d,} Profile of the magnetization along the domain wall 
projected along S$_z$ and the in-plane (S$_x$, S$_y$) components, respectively. 
Fitting curves are obtained using Eqs. \ref{eq:Sxy}$-$\ref{eq:Sz}. 
We computed domain wall widths $\sigma_z=3.41$ nm 
($\pm 0.03$) and $\sigma_x,y=3.50$ nm 
($\pm 0.06$). 
{\bf e-f,} Top and side views, respectively, of the rotation 
of the magnetization along the domain wall.  
Both S$_x$ and S$_y$ show variations along 
the wall altogether with S$_z$ which indicate 
a hybrid character of the domain wall, 
i.e. neither Bloch nor N\'eel.  Colours follow the scale bar in {\bf b}.
}
\end{figure}

\begin{figure}[htbp]
\centering
\caption{\label{fig3}\textbf{Field- and current-induced domain wall dynamics 
in a 2D antiferromagnet}  
{\bf a,} Snapshots of the domain wall dynamics in MnPS$_3$ induced by 
electric currents $j$(10$^{9}$ A cm$^{-2}$) and magnetic fields B(T). 
The initial domain wall configuration ($j$=0, B=0) 
is the same for both driving forces at t=0 ns (middle panel).  
The current-driven domain wall motion is shown at different current 
densities ($2 \times 10^{8} - 5 \times 10^{10}$ A cm$^{-2}$) 
but at the same time evolution of 0.5 ns (upper panels). 
For the field-induced domain wall motion (lower panels), 
two magnitudes at 2 T and 4 T are shown.  
A 300 nm$\times$50 nm flake is considered in all simulations.
{\bf b,} Simulated wall velocity $v$ (m s$^{-1}$) versus current 
density $j$ (10$^{9}$ A cm$^{-2}$) (triangles)
considering two fits to the data. In the low-velocity regime, 
Eq.\ref{eq:linear_v} is used to describe the linear dependence (black curve). 
In the relativistic regime,
Eq.\ref{eq:relativ}  
provides an accurate description 
over the entire range of densities. Maximum group velocities 
$v_{g1}$ and $v_{g2}$ are shown for comparison 
via horizontal dashed lines in the coloured region.  
{\bf c,} Variation of the out-of-plane component of the magnetization S$_{\rm z}$ 
across the domain wall with a current density of 
2 $\times$10$^{9}$ A cm$^{-2}$ at t=0.0, 0.25, 0.5 ns. 
The calculated points are fitted to 
Eq.\ref{eq:Sz} shown with the solid line. The variation in position 
of the centre of the domain wall as a function of time is used to 
extract the wall velocity which is an average over all atoms at the 
domain wall. 
{\bf d-e,} Current-driven domain wall width $\sigma$ (nm) and 
domain wall mass $M_{\rm DW}$, respectively, 
versus $v$. Fits to Eq. \ref{eq:width} and Eq. \ref{eq:mass} are 
included for comparison.  
{\bf f-g,} In-plane components of the magnetization 
S$_{x,y}$ (a.u.) versus time (ps), respectively, 
for $j=2 \times 10^{9}$ A cm$^{-2}$. 
Frequencies within the range of 0.79$-$0.81 THz can be extracted 
from S$_{x,y}$ shown via the solid curves in a full circle. 
{\bf h,} Close look of the snapshot of the 
field-induced domain wall motion at B=2.0 T in {\bf a}. 
The presence of different edges (zigzag and dangling-bond) 
terminating the layer along $y$
induces a bending of the wall profile under the field 
and consequently a slight asymmetry in the 
displacement of the wall. Only Mn atoms in the 
honeycomb lattice are shown. 
{\bf i,} Wall velocity versus magnetic field applied perpendicular to the surface. 
The dashed line is given by $y= 86.28 x$ with linear regression coefficient R$^2=0.9996$.    
{\bf j,} Similar as {\bf c} but at an applied field of 2.0 T. The initial condition shows 
that the spin directions of all atoms at the same $x$ (and any $y$)
appear superimposed on top of each other, revealing a highly ordered system. 
As the wall motion starts, atoms at the same $x$ but different $y$ 
no longer have the same spin direction, leading to a continuous distribution of spins. 
} 
\end{figure}

\begin{figure}[htbp]
\centering
\caption{\label{fig4}\textbf{Sublattice mediated domain-wall motion in a 2D anti-ferromagnet under magnetic fields.} 
{\bf a,} Schematic of the intended wall motion taking place in opposite directions 
at different sublattices (i.e. A or B) under an 
external field B$_{\rm z}$. Only atoms at one sublattice (green or violet) are
shown around each edge to facilitate the view. 
The rotation of the spins over time at sublattices A (green) and B (violet) are represented 
with the arrows changing systematically 
in the background. Only Mn atoms are shown. 
{\bf b-c,} Monolayer MnPS$_3$ terminated with both edges   
in zig-zag (ZZ) configuration, and with a combination of ZZ 
and dangling-bond (DB) arrangements, respectively. 
The domain wall is shown at the faint atoms in the middle of the layer. 
{\bf d,} Diagram of the energy versus the angle $\theta$ defined relative to the $z$-axis. 
Away from the domain wall, $\theta$ can be either $0^{\rm o}$ or $180^{\rm o}$ 
depending on what sublattice is considered. 
At the wall, $\theta$ can range within $0-180^{\rm o}$ for one sublattice, and 
$180-0^{\rm o}$ for the other. B$_{\rm z}$ points along of $z<0$. 
Spins at the wall ($0^{\rm o}<\theta<180^{\rm o}$, 
which excludes fully spin-up and spin-down states) 
react differently than those away from the wall 
(i.e., $\theta=0$ for spin-up or $theta=180^{\rm o}$ 
for spin down) to an external magnetic field. Only for those at the wall,
a finite B$_{\rm z}$ changes the energetic stability of the system  
inducing a rotation of the spins as the magnitude of 
$\theta$ changes to a new energy minimum 
(e.g. $\theta_{B_z=0} \ne \theta_{B_z \ne 0} $). 
$\Delta{\rm E}$ shows the energy gained through a rotation to 
the new minimum once the field is applied. 
For the spins away from the wall, B$_{\rm z}$ causes 
a rigid shift of the energy curve while preserving its shape. 
This results in no change in the value of $\theta$ 
for the minimum energy, and thus, no rotation 
induced locally by the magnetic field.
The energy is calculated via Eq. \ref{BfieldHamilton}.  
{\bf e-f,} Plots of $-\Delta{\rm E}$ for few atoms at 
different regions of the layer such as at the edges, near the 
edges and middle of the sheet for systems with ZZ-ZZ and ZZ-DB 
edges, respectively. The different sublattices (A and B) 
are shown individually in different coloured curves. 
We plot $-\Delta{\rm E}$ instead of $\Delta{\rm E}$ to 
better display the variations of energy at different parts of the system.  
The inset in {\bf f} shows a side view of the layer with the 
dimensions considered in the model. 
}
\end{figure}

\pagebreak{}

\pagebreak 


\setcounter{figure}{0}

\begin{figure}[htbp]
\centering
\includegraphics[width=1.05\linewidth]{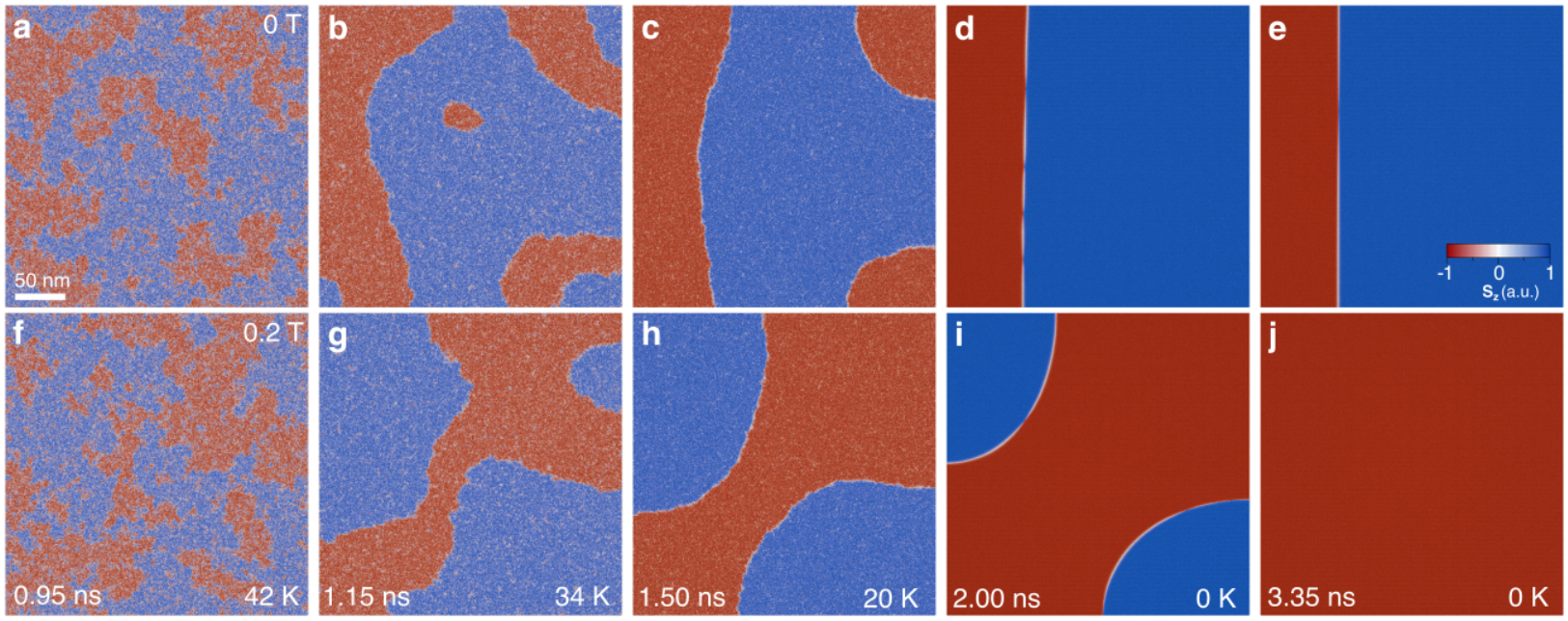} 
\caption{}
\end{figure}

\begin{figure}[htbp]
\centering
\includegraphics[width=1.05\linewidth]{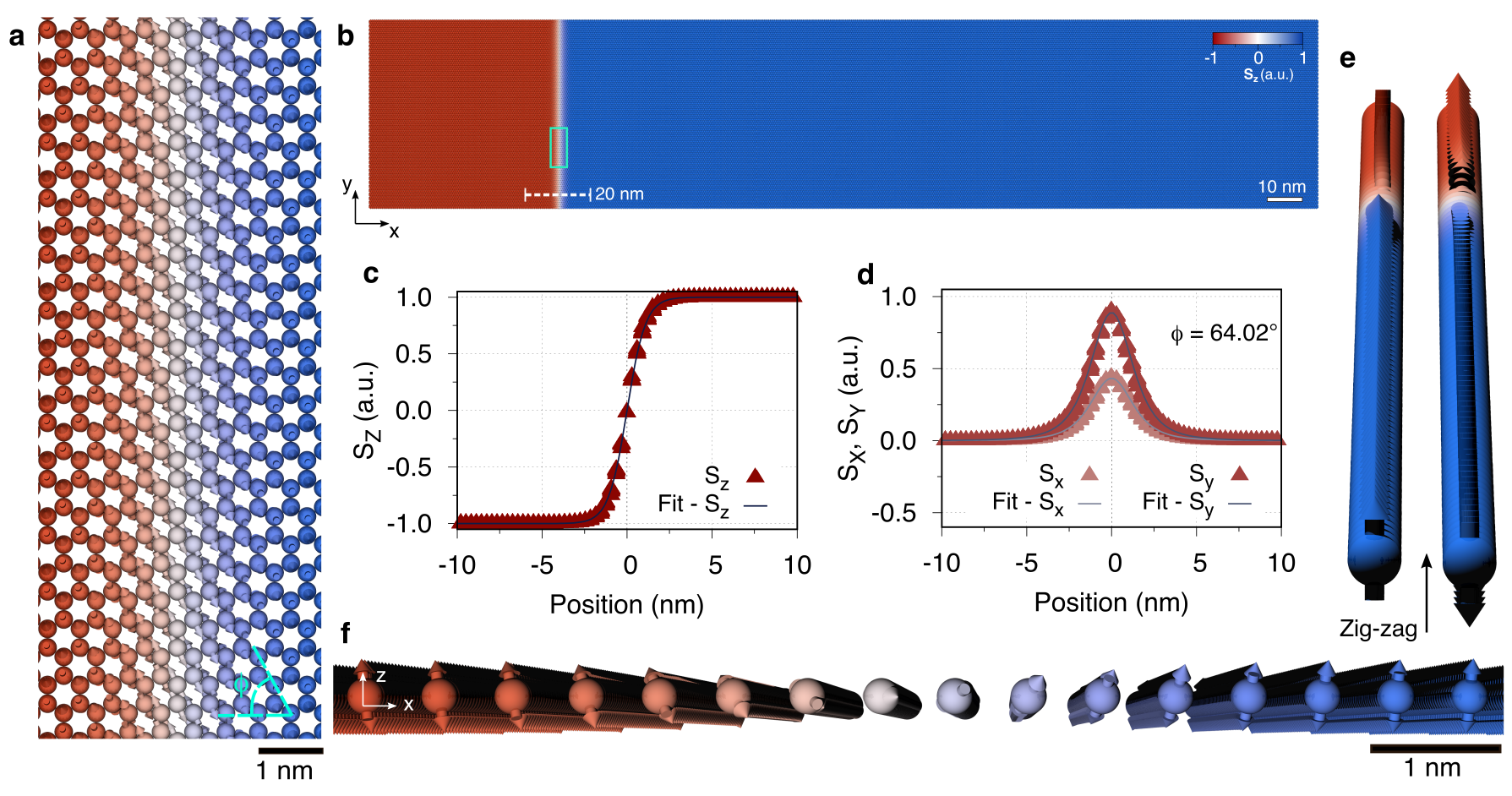} 
\caption{}
\end{figure}



\begin{figure}[htbp]
\centering
\includegraphics[width=1.10\linewidth]{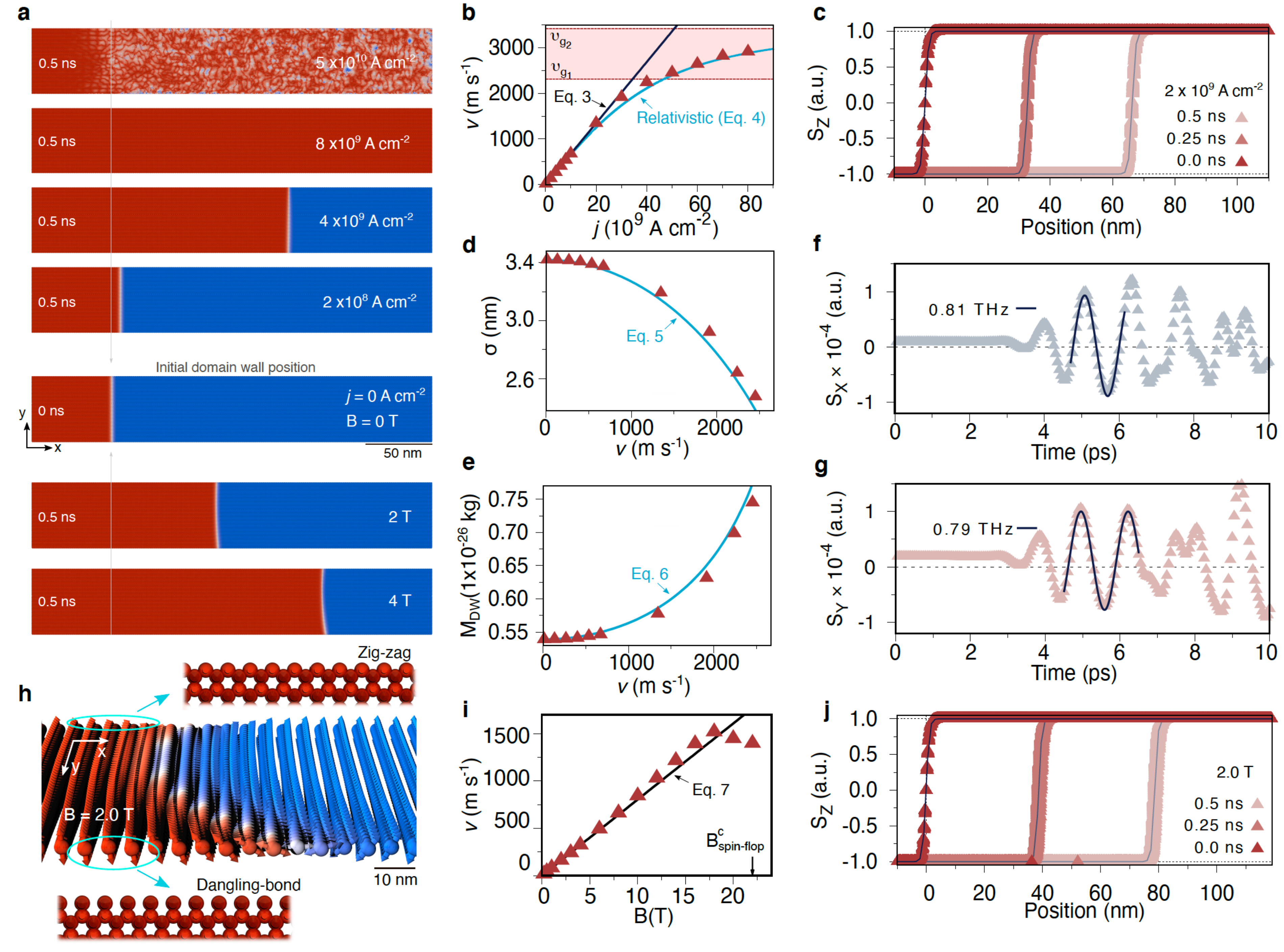}
\caption{}
\end{figure}

\begin{figure}[htbp]
\centering
\includegraphics[width=1.18\linewidth]{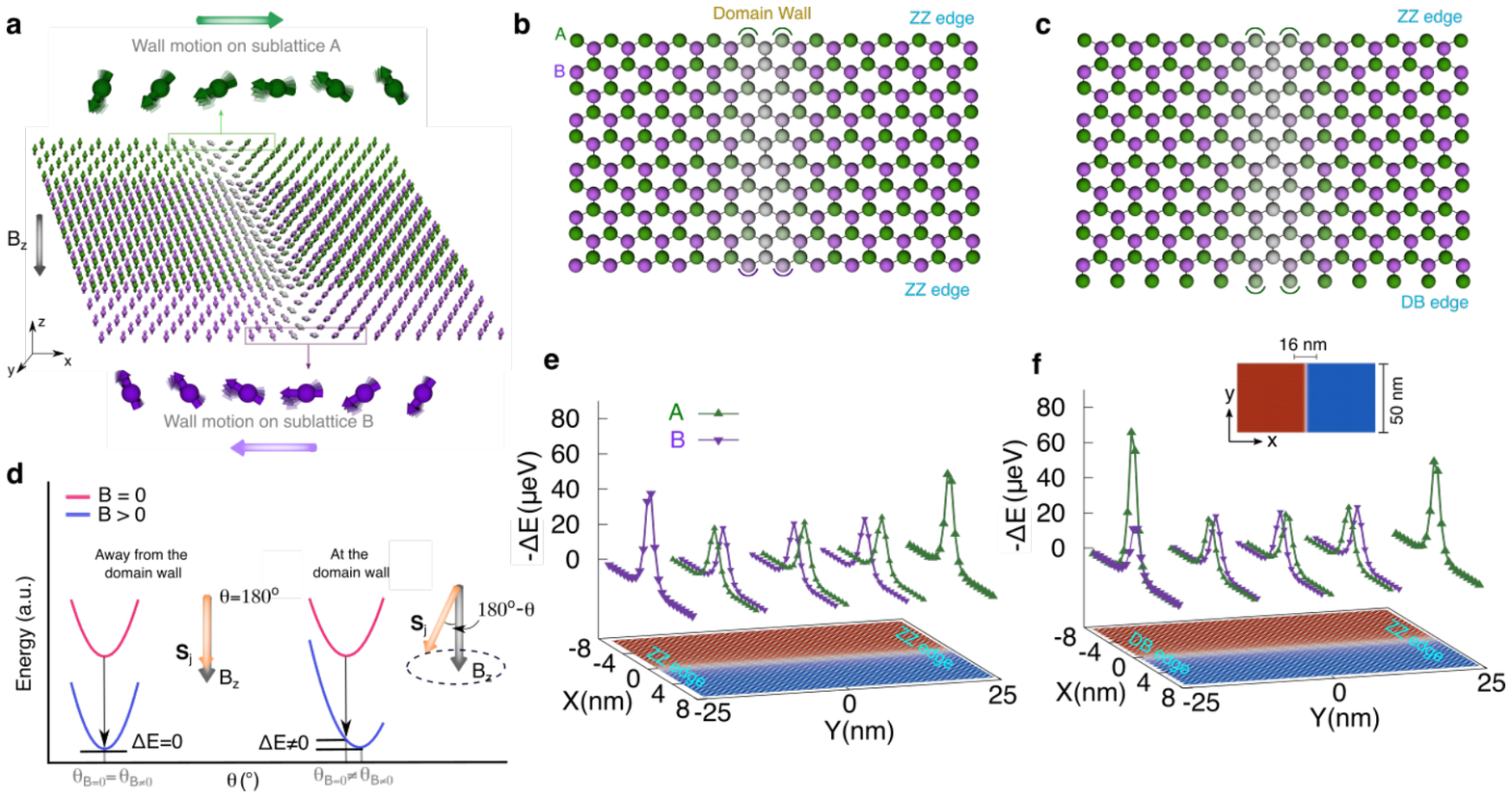} 
\caption{}
\end{figure}

\pagebreak{}


%

\end{document}